\documentclass[11pt]{article}
\usepackage{amsmath,amsfonts,amssymb,bm}
\RequirePackage[a4paper,body={16cm,26cm}]{geometry}
\usepackage{pslatex}
\usepackage{graphicx,color}
\usepackage{etex}
\usepackage[T1]{fontenc}
\usepackage[utf8]{inputenc}
\usepackage[french]{babel}
\usepackage[noindentafter,pagestyles]{titlesec}
\usepackage{textcomp}
\usepackage{dcolumn}
\usepackage{amsmath}
\usepackage{txfonts}
\usepackage{fancyhdr}
\usepackage{threeparttable}
\usepackage{xspace}
\usepackage[superscript]{cite}
\usepackage{subcaption}
\usepackage{pictex}

\usepackage[algo2e,ruled,lined,titlenumbered,commentsnumbered]{algorithm2e}
\usepackage[export]{adjustbox}
\usepackage{graphicx,color}
\usepackage{algpseudocode}
\usepackage{algorithm2e}
\usepackage{algorithm}
\usepackage{authblk}
\usepackage{hyperref}
\usepackage{float}
\newcommand{\address}[1]{\def\@address{#1}}
\title{Couplage Global-Local en asynchrone pour des problèmes linéaires}
\author{Ahmed EL KERIM$^{1,3}$, Pierre GOSSELET$^2$, Frédéric MAGOUL\`ES$^3$, \\}
\date{$^1$ Universit\'e Paris-Saclay, ENS Paris-Saclay, CNRS, LMT, Gif-sur-Yvette, France, ahmed.elkerim@ens-paris-saclay.fr\\
	$^2$ Université de Lille, CNRS, Centrale Lille / LaMcube, pierre.gosselet@univ-lille.fr\\
	$^3$ Universit\'e Paris-Saclay, CentraleSup\'elec / MICS , Gif-sur-Yvette, France, frederic.magoules@hotmail.com\\
}
\begin{document}
		\maketitle
		\newcommand{\dom}{\ensuremath{\Omega}}
		\newcommand{\inter}{\ensuremath{\Gamma}}
		\newcommand{\F}{\ensuremath{^F}}
		\newcommand{\G}{\ensuremath{^G}}
		\newcommand{\GT}{\ensuremath{^{G^T}}}
		\newcommand{\Fs}{\ensuremath{^{(s),F}}}
		\newcommand{\Gs}{\ensuremath{^{(s),G}}}
		\newcommand{\Gz}{\ensuremath{^{(0),G}}}
		\newcommand{\s}{\ensuremath{^{(s)}}}
		\newcommand{\z}{\ensuremath{^{(0)}}}
		\newcommand{\sT}{\ensuremath{^{(s)^T}}}
		\newcommand{\zT}{\ensuremath{^{(0)^T}}}
		\newcommand{\R}{\ensuremath{^R}}
		\newcommand{\Gm}{\ensuremath{^{G^{-1}}}}
		\newcommand{\bK}{\ensuremath{\mathbf{K}}}
		\newcommand{\bA}{\ensuremath{\mathbf{A}}}
		\newcommand{\bI}{\ensuremath{\mathbf{I}}}
		\newcommand{\bJ}{\ensuremath{\mathbf{J}}}
		\newcommand{\bS}{\ensuremath{\mathbf{S}}}
		\newcommand{\bT}{\ensuremath{\mathbf{T}}}
		\newcommand{\bP}{\ensuremath{\mathbf{P}}}
		\newcommand{\bM}{\ensuremath{\mathbf{M}}}
		\newcommand{\bSnl}{\ensuremath{\mathcal{S}}}
		\newcommand{\bu}{\ensuremath{\mathbf{u}}}
		\newcommand{\bp}{\ensuremath{\mathbf{p}}}
		\newcommand{\bq}{\ensuremath{\mathbf{q}}}
		\newcommand{\bv}{\ensuremath{\mathbf{v}}}
		\newcommand{\br}{\ensuremath{\mathbf{r}}}
		\newcommand{\bb}{\ensuremath{\mathbf{b}}}
		\newcommand{\bx}{\ensuremath{\mathbf{x}}}
		\newcommand{\f}{\ensuremath{\mathbf{f}}}
		\newcommand{\foint}{\ensuremath{\mathbf{f}_{int}}}
		\newcommand{\foext}{\ensuremath{\mathbf{f}_{ext}}}
		\newcommand{\foexti}{\ensuremath{\mathbf{f}_{ext,i}}}
		\newcommand{\foextb}{\ensuremath{\mathbf{f}_{ext,b}}}
		\newcommand{\lam}{\ensuremath{\boldsymbol{\lambda}}}
		\begin{abstract}
			Une version parallèle asynchrone du couplage global-local non-intrusif est mise en place. Le cas de nombreux patchs, y compris couvrant l'intégralité de la structure est étudié. L'asynchronisme permet de limiter la dépendance aux communications, aux pannes et au déséquilibre de charge. Nous détaillons la méthode et illustrons ses performances sur un cas académique.

			An asynchronous parallel version of the non-intrusive global-local coupling is implemented. The case of many patches, including those covering the entire structure, is studied. The asynchronism limits the dependency on communications, failures, and load imbalance. We detail the method and illustrate its performance in an academic case.
			
			\textbf{Mot clés:}{calcul parallèle asynchrone, couplage non-intrusif, décomposition de domaine}

		\end{abstract}
	%\keywords calcul parallèle asynchrone, couplage non-intrusif, décomposition de domaine.
	\section{Introduction}
	De nombreuses modélisations industrielles sont définies par une hiérarchie de modèles. À chaque niveau, le modèle le plus grossier permet d'estimer les grands flux d'efforts dans le domaine, il est enrichi par un ensemble de patchs fins qui viennent préciser localement la géométrie, le maillage, les propriétés matériau. Pour coupler les modèles, la stratégie couramment utilisée est le zoom structural (\textit{submodeling}) qui consiste à imposer des conditions de Dirichlet issues du calcul grossier sur les modèles fins. En négligeant les effets des patchs sur le modèle global, la méthode est susceptible de conduire à de grandes erreurs sur les quantités mécaniques d'intérêt.
	
	Le couplage global/local \cite{Gendre} consiste à faire remonter l'influence des patchs par un effort appliqué au modèle global. Cet effort est obtenu à l'aide d'itérations qui permettent une mise en œuvre non-intrusive de la méthode autour de codes commerciaux (par exemple code\_aster ou abaqus). Cette méthode peut s'interpréter de nombreuses manières, notamment comme une méthode de décomposition de domaine de Schwarz \cite{Hecht,Gosselet}. 
	
	Ce genre de méthode est bien adapté au calcul parallèle, et des travaux récents \cite{Magoules1, Garay, Magoules2, Glusa, Guillaume} ont montré qu'elles supportaient les itérations asynchrones qui permettent une meilleure tolérance aux latences réseaux, au déséquilibre de charge et aux architectures fortement hétérogènes.
	
	Ce papier présente la version asynchrone du couplage global/local. Après une présentation rapide des fondamentaux de la méthode dans la section~\ref{sec:gl}, l'asynchronisme est introduit dans la section~\ref{sec:async}. La section~\ref{sec:valid} présente un étude numérique de validation sur un cas-test académique d'élasticité linéaire 3D.
	
\section{Couplage Global-Local}\label{sec:gl}
\begin{figure}[hbt!]
	\centering\null\hfill
	\begin{minipage}{0.3\textwidth}
		\includegraphics[width=\textwidth]{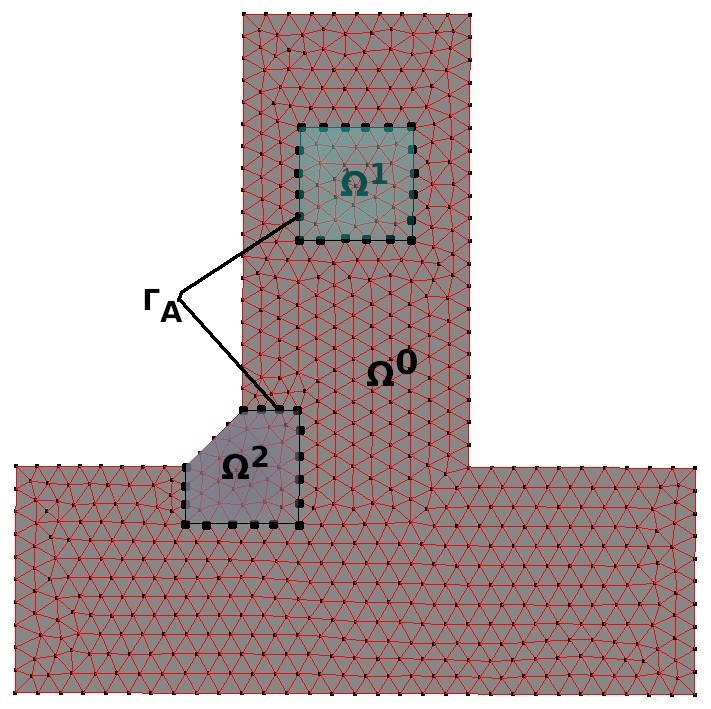}
		\caption{Problème Global}\label{fig:Glo}
	\end{minipage}
	\hfill
	\begin{minipage}{0.3\textwidth}
		\includegraphics[width=\textwidth]{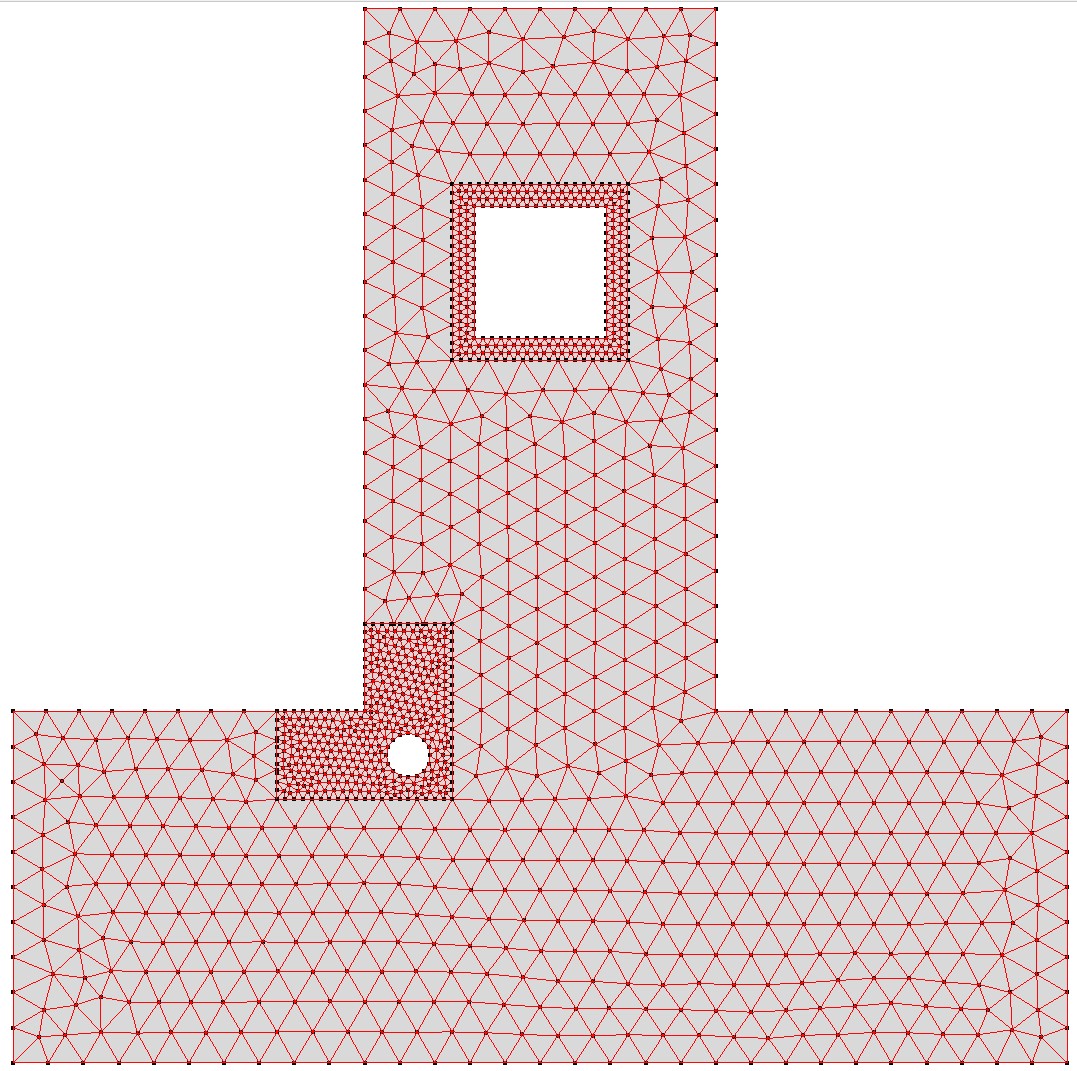}
		\caption{Référence}\label{fig:Ref}
	\end{minipage}\hfill\null
\end{figure}

On considère une structure définie à plusieurs échelles : un modèle Grossier représentant la globalité du domaine est localement corrigé par $N$ patchs Fins. Les patchs peuvent couvrir l'entièreté du domaine \cite{Duval} ou non \cite{Allix}. On note $\Omega^{(s)}$ avec $s>0$ les zones d'intérêt dont il existe une représentation Grossière et une représentation Fine. Si elle existe, on note $\Omega^{(0),G}$ la zone Grossière non couverte par des patchs, parfois appelée zone complémentaire. L'ensemble $(\Omega^{(s),G})_{s\geqslant0}$ constitue le modèle Grossier, alors que $(\Omega^{(0),G},(\Omega^{(s),F})_{s>0})$ constitue la référence, comme illustré sur les figures~\ref{fig:Glo} et~\ref{fig:Ref}.

L'objectif du couplage est de trouver la solution du problème de référence en alternant les calculs sur le modèle grossier et sur les modèles fins.

\subsection{Problème global et problème de référence}
On se place dans des hypothèses de thermique ou mécanique linéaire (élasticité en petite perturbation). On utilise une discrétisation élément fini. On note $\bK$ les matrices de rigidité (ou leur équivalent thermique) qui sont symétriques semi-définies positives, $\foext$ les vecteurs de chargements imposés, et $\bu$ le vecteur inconnu (température ou déplacement).

Le problème Global, indicé par $G$, est une simplification du problème, permettant un calcul rapide mais représentatif des flux à grande échelle. Il s'écrit :
%Comme precisé avant on definit ici le problème global indexé par $G$ comme une simplification du problème, adapté à un calcul rapide et capable de donner une représentation correcte des flux à longue distance:
\begin{equation} 
	-\bK\G\bu\G + \foext\G =0.
\end{equation}
Suite à la décomposition en sous-domaines (complémentaire et patchs), on peut séparer les degrés de liberté de l'interface (indexé par $b$) et les degrés de liberté internes (indexé par $i$). On peut réécrire l'équilibre grossier sur chaque sous-domaine en introduisant les réactions nodales $\lambda\Gs$ et l'opérateur de trace $\bT\s$ tel que  $\bT\s\bu\Gs=\bu\Gs_b$ :
%Pour simplifier la manipulation on divise le problème global en $N+1$ sous-domaines sans recouvrement $(\Omega\s)$ avec $s\in[0..N]$, les sous-domaines donc se coincident à l'interface. Après une discrétisation classique par éléments finis, on definit l'operateur de la trace sachant que $\bT\s\bu\Gs=\bu\Gs_b$ \\ et on introduit les reactions nodales  $\lambda\Gs$ sur l'interface. Le problème à résoudre pour chaque sous-domaine peut être écrit comme suit :
\begin{equation}\label{eq:FEbalance}
	-\bK\Gs\bu\Gs + \foext\Gs + \bT\sT \lam\Gs=0.
\end{equation}
Ce système s'écrit sous forme détaillée :
\begin{equation}
	\begin{pmatrix}
		\bK\Gs_{ii} & \bK\Gs_{ib} \\
		\bK\Gs_{bi} & \bK\Gs_{bb} \\
	\end{pmatrix}
	\begin{pmatrix}
		\bu\Gs_{i}\\
		\bu\Gs_{b}\\
	\end{pmatrix}= 	\begin{pmatrix}
		\foexti\Gs\\
		\foextb\Gs + \lam\Gs\\
	\end{pmatrix}.
\end{equation}
D'après les hypothèses, nous pouvons éliminer les degrés de liberté internes, condenser le problème sur l'interface et introduire un opérateur de Dirichlet-Neumann $\bS$ (complément de Schur primal) :
\begin{equation}
	\lam\Gs = \bS\Gs \bu\Gs_b - \bb\Gs \quad\text{avec} \quad \left\{
	\begin{aligned}
		& \bS\Gs = \bK\Gs_{bb} - \bK\Gs_{bi} (\bK\Gs_{ii})^{-1} \bK\Gs_{ib}  \\
		& \bb\Gs = \foextb\Gs - \bK\Gs_{bi} (\bK\Gs_{ii})^{-1} \foexti\Gs
	\end{aligned}\right..
\end{equation}
Afin de connecter les sous-domaines, nous introduisons les opérateurs d'assemblage $(\bA\s)$, qui sont des matrices booléennes creuses qui projettent les bords des sous-domaines sur l'interface globale. Les conditions d'interface peuvent être écrites comme suit :
\begin{itemize}
	\item continuité du déplacement : $\exists \bu_A\G$ tel que $\forall s\in[0..N],\ \bu_b\Gs=\bA\sT\bu_A\G$,
	\item équilibre des réactions nodales : $\sum_{s=0}^N \bA\s \lam\Gs = 0$.
\end{itemize}
Avec ces notations, le problème global s'écrit :
\begin{equation}\label{eq:global}
	\begin{aligned}
		&	\text{Trouver le déplacement global } \bu\G_A \text{ tel que}\\
		&	\underset{\bS\G}{\underbrace{\left(\sum_{s=0}^N \bA\s \bS\Gs\bA\sT\right)}}\bu\G_A  = \underset{\bb\G}{\underbrace{\sum_{s=0}^N \bA\s \bb\Gs}}
	\end{aligned}
\end{equation}
Il suffit de supposer qu'au moins un des sous-domaines possède suffisamment de conditions de Dirichlet imposées pour que ce problème soit symétrique défini positif et donc bien posé.

Le problème de référence consiste à remplacer les zones d'intérêt grossières par leur représentation fine. De même que pour le grossier, on utilise la condensation pour faire apparaître les opérateurs d'interface. Pour plus de souplesse, on autorise des maillages grossiers et fins non  aux interfaces, et on introduit l'opérateur d'interpolation entre le global et le local, noté $\bJ^(s)$. On adopte la cinématique grossière à l'interface, de sorte que $\bu_b\Fs=\bJ\s\bu_b\Gs$. De même l'équilibre des réactions est évalué sur l'interface globale. 

Ces hypothèses conduisent à formuler le problème de référence sous la forme suivante :
\begin{equation}\label{eq:reference}
	\begin{aligned}
		&		\text{Trouver le déplacement de référence } \bu\R_A \text{ tel que}\\
		&	\underset{\bS\R}{\underbrace{\left(	\bA\z \bS\Gz\bA\zT + \sum_{s=1}^N \bA\s\bJ\sT \bS\Fs\bJ\s\bA\sT\right)}} \bu\R_A  = \underset{\bb\R}{\underbrace{\bA\z \bb\z + \sum_{s=1}^N \bA\s \bJ\sT\bb\Fs}}.
	\end{aligned}
\end{equation}

\subsection{Le couplage}
Le couplage global-local est une technique non-intrusive qui permet de retrouver la solution du problème de référence en résolvant des problèmes grossiers modifiés par l'introduction d'un effort supplémentaire (opération supportée par tous les codes, même commerciaux) et en résolvant des problèmes fins indépendants par patchs.

On peut interpréter l'algorithme comme la résolution de \eqref{eq:reference} en utilisant un préconditionnement à droite, autrement dit un changement de variable. On introduit une charge externe $\bp_A\G$ appliquée à l'interface $\Gamma$ dans le problème global et on exprime le déplacement de référence comme la solution du problème global modifié :
%sachant que le problème soit en equilibre sur cette interface. La procedure iterative consiste donc à corriger la valeur de $\bp_A\G$ en connaissant la difference entre les reaction nodales sur l'interface du modèle local et du modèle global. Le problème global dans ce cas s'écrit comme :
\begin{equation}\label{eq:globalwithp}
	\begin{aligned}
		&	\text{Définir le déplacement de référence }\bu\R_A \text{ tel que }\\
		&	\bS\G \bu\R_A = \bp_A\G + \bb\G
	\end{aligned}
\end{equation}
On a $\bp_A\G = \sum \bA\s \lam\Gs$ et donc $\bp_A\G$ peut être interprété comme un manque d'équilibre imposé aux réactions nodales sur l'interface dans le domaine global. Ce déséquilibre dans le modèle global doit compenser son inacuité par rapport au modèle fin :
\begin{equation}\label{eq:refwithp}
	\begin{aligned}
		&\text{Trouver l'intereffort }\bp\G_A \text{ tel que }\\
		&\bS\R \bS\Gm\bp_A\G  = \bb\R- \bS\R \bS\Gm\bb\G.
	\end{aligned}
\end{equation}

Si on utilise $\bS\R=(\bS\R-\bS\G)+\bS\G$, on peut reécrire le problème sous la forme suivante, où on introduit le résidu $\br$ :
\begin{equation}\label{eq:globallocalb}
	\begin{aligned}
		\text{Trouver }\bp_A\G \text{ tel que }
		\br := -\left(\bp_A\G	- (\bS\G-\bS\R)\bS\Gm\bp_A\G - (\bb\R- \bS\R \bS\Gm\bb\G)\right) =0
	\end{aligned}
\end{equation}
Cette équation invite à utiliser un algorithme itératif stationnaire, le plus simple étant la méthode de Richardson $\bp_{A,j+1}\G=\bp_{A,j}\G+\br_{j}$, mais de nombreuses techniques d'accélération sont possibles et ont été testées avec succès comme Aitken, Quasi-Newton, gradient conjugué (voir \cite{Gosselet} pour une rapide revue).

Il est à noter que le résidu possède une interprétation mécanique simple. Partant d'un intereffort $\bp_{A,j}$, on calcule le déplacement global $\bu_{A,j}$ et l'éventuelle réaction de la zone complémentaire $\lam\Gz_j$ ; on applique $\bJ\s\bu_{A,j}$ sur les modèles fins et on récupère la réaction $\lam\Fs_j$ associée ; le résidu correspond à :
\begin{equation}\label{eq:residwith0}
	\br_j = - \left(\bA\z\lam\Gz_j + \sum_{s=1}^N\bA\s\bJ\sT\lam\Fs_j\right).
\end{equation}
Autrement dit, le résidu correspond au déséquilibre des efforts de réaction entre les sous-domaines du problème de référence.

L'algorithme~\ref{alg:syncAit} présente la version synchrone du couplage global-local. Le vocabulaire du parallélisme avec échange de messages est employé.
\begin{algorithm}\caption{Algorithme du couplage synchrone avec relaxation }\label{alg:syncAit}\DontPrintSemicolon
	Initialisation de $\bp_A\G=0$ et $(\bq\s=0)$ pour le model \textit{global} (rank 0)\;
	\If{Rang 0}{
		Réception de tous les $\bq\s$ des autres rangs\;
		Calcul du résidu: $\br_A= - \sum_{s=0}^N\bA\s\bq\s$\; 
		\If{$\|\br_A\|$ est suffisamment petit (initialisation exclue)}{break}
		Màj $\bp_A =\bp_A + \omega\br_A$\;
		Résolution globale: $\bu\G_{A}= \bS\Gm(\bb\G+\bp_{A}\G)$\;
		Envoi de $(\bA\sT\bu\G_A)$ sur les rangs locaux \;
		S'il y a un complémentaire, calcul de $\bq\z:=\lam\Gz$.
	}
	\If{Rang $s>0$ (patchs)}{
		Réception de $\bu_b\Fs:=\bJ\s\bA\sT\bu_A\G$ du Rang 0\;
		Résolution locale fine $\lam\Fs = \bS\Fs \bu\Fs_b - \bb\Fs$\;
		Envoi de $\bq\s:=\bJ\sT\lam\Fs$ au Rang 0\;
	}
\end{algorithm}
\section{Version Asynchrone}\label{sec:async}
En asynchrone, on doit présumer que les données disponibles peuvent être en retard par rapport à l'itération courante. Dans la méthode globale-locale, le calcul global joue un rôle centralisateur, et on introduit les fonctions ${\sigma\s(j)}\leqslant j$ pour symboliser le retard dans l'information issue du sous-domaine $s$ disponible à l'itération $j$. On reprend en asynchrone le schéma itératif de Richardson en employant l'équation~\eqref{residwith0} pour calculer le résidu avec les informations disponibles :
\begin{equation}\label{eq:residwith0}
	\br_j = - \bA\z\lam\Gz_j - \sum_{s=1}^N\bA\s\bJ\sT\lam\Fs_{\sigma\s(j)}
\end{equation}
Pour illustrer les approches synchrone et asynchrone, nous utilisons le problème de la figure~\ref{fig:Ref} avec 2 patchs. Le graphique~\ref{fig:sync} montre une séquence d'itérations dans le cas synchrone du couplage global-local où nous pouvons voir l'effet de la synchronisation et le temps d'attente nécessaire à chaque processeur avant de passer d'une itération à une autre en attendant de recevoir ou d'envoyer des informations :
\begin{figure}[ht]
	\centering
	\begin{minipage}{.49\textwidth}
		\includegraphics[width=.9\textwidth]{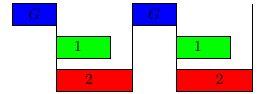}
		\caption{Deux itérations synchrones}
		\label{fig:sync}
	\end{minipage}
	\begin{minipage}{.49\textwidth}
		\includegraphics[width=.55\textwidth]{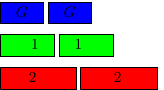}
		\caption{Deux \og{}itérations asynchrones\fg{}}
		\label{fig:async}	
	\end{minipage}
\end{figure}

Dans le graphique~\ref{fig:async} et l'algorithme~\ref{alg:asyncAit}, nous observons la séquence d'itérations dans le cas du couplage asynchrone où chaque processeur avance avec les données disponibles pour continuer, et rend son information aux autres dès qu'il a fini de la calculer. L'algorithme~\ref{alg:asyncAit} utilise le vocabulaire du parallélisme par accès mémoire à distance (RMA).

\begin{algorithm}\caption{Algorithme du couplage asynchrone}\label{alg:asyncAit}\DontPrintSemicolon
	Initialisation $\bp_A\G=0$ et  $(\bq\s=0)$ sur le modèle \textit{global} (rank 0)\;
	\If{Rang 0 détecte la réception d'une nouvelle valeur $\bq\s$ d'un rang $s$} {
		Calcul du residu: $\br_A= - \sum_{s=0}^N\bA\s\bq\s$\; 
		\If{$\|\br_A\|$ est suffisament petit (initialisation exclue)}{break}
		Màj $\bp_A =\bp_A + \omega\br_A$\;
		Résolution globale: $\bu\G_{A}= \bS\Gm(\bb\G+\bp_{A}\G)$\;
		Global \textit{puts} $(\bA\sT\bu\G_A)$ sur les rangs locaux \;
		S'il y a un complémentaire, calcul de $\bq\z:=\lam\Gz$.
	}
	\If{sous-domaine $s>0$ est disponible et détecte l'arrivée d'une nouvelle valeur $\bu\Fs_b:=\bJ\s\bA\sT\bu_A\G$}{
		Résolution locale fine $\lam\Fs = \bS\Fs \bu\Fs_b - \bb\Fs$\;
		Patch $s$ \textit{puts} $\bq\s:=\bJ\sT\lam\Fs$ sur le Rang 0\;
	}
\end{algorithm} 

La détection de convergence peut se révéler complexe en asynchrone \cite{Magoules4, Miellou}. Un grand avantage du couplage global/local est l'assemblage du résidu sur le modèle grossier, il est donc très simple de calculer un critère d'arrêt. Concernant la preuve théorique de la convergence de la méthode, l'utilisation du cadre des paracontractions \cite{Eisner} permet de démontrer que si un coefficient de relaxation $\omega$ permet de faire converger la méthode synchrone, alors ce même coefficient est suffisant pour faire converger en asynchrone.

Du point de vue de l'implémentation, plusieurs techniques existent dans la littérature. Une première repose sur des communications MPI point-à-point non bloquantes comme dans \cite{Magoules3}. Une seconde, comme \cite{Yamazaki}, se base sur les communications unilatérales ou MPI- RDMA, qui consistent à autoriser un processeur à directement lire ou écrire le contenu de la mémoire des autres processeurs. Nous avons retenu cette deuxième stratégie dans notre mise en œuvre.

\subsection{Validation sur un cas académique}\label{sec:valid}

\begin{figure}[htb]
	\begin{minipage}[b]{0.3\linewidth}
		\centering
		\includegraphics[scale=0.3]{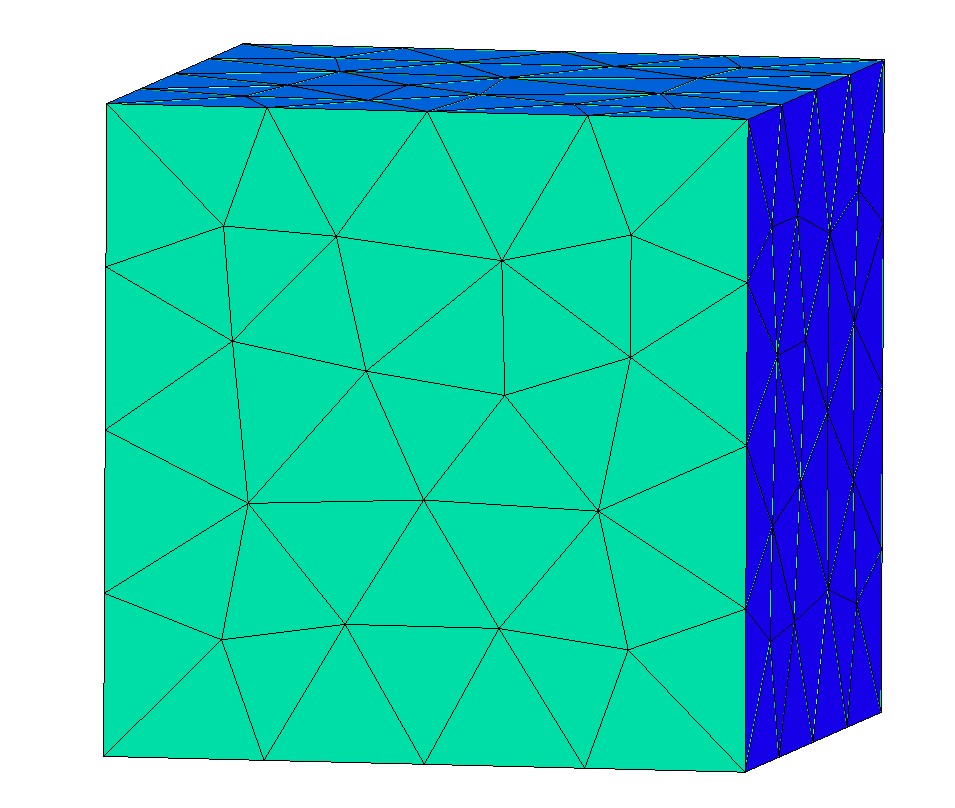}
		\caption{Global $\Omega\Gs$}\label{fig:exglo}
	\end{minipage}
	\begin{minipage}[b]{0.3\linewidth}
		\centering
		\includegraphics[scale=0.3]{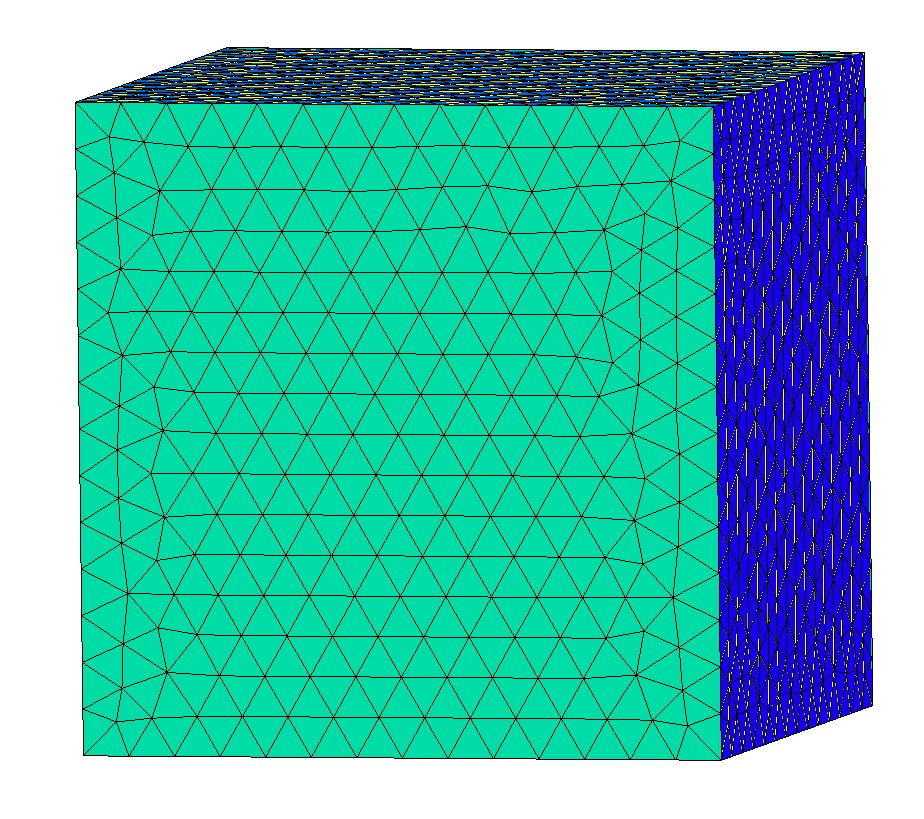}
		\caption{Fin $\Omega\Fs$}\label{fig:exfin}
	\end{minipage}
	\begin{minipage}[b]{0.3\linewidth}
		\centering
		\includegraphics[scale=0.4]{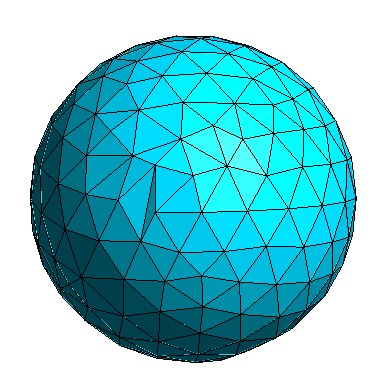}
		\caption{$\Omega\Fs_{in}$: Inclusion dans le fin $\Omega\Fs$}\label{fig:exfinI}
	\end{minipage}
\end{figure}

Pour réaliser notre étude numérique, nous avons construit un cas test permettant d'évaluer l'extensibilité faible de la méthode. Il s'agit d'une \og{}poutre 3D\fg{} hétérogène en élasticité linéaire, constituée de la répétition selon les 3 axes d'un sous-domaine cubique. La version fine du sous-domaine comporte une inclusion sphérique de module d'Young 10 fois plus faible que celui de la matrice, alors que la version grossière est homogène constitué uniquement de matrice (le coefficient de Poisson est constant, égal à 0,3), avec un maillage moins fin voir les figures~\ref{fig:exglo},~\ref{fig:exfin} et~\ref{fig:exfinI}. Pour ce cas test, les patchs fins recouvrent l'intégralité du domaine, il n'y a donc pas de zone complémentaire $\Omega\z$. Le chargement volumique est identique dans tous les modèles $f=(1,1,1)^T$. Un côté de la poutre est encastré, les autres faces sont libres.

La géométrie est générée à l'aide du logiciel gmsh \cite{Geuzaine} et la bibliothèque getfem \cite{Renard} est utilisée depuis notre code python pour mettre en place l'approximation élément fini. Les calculs ont été conduits sur le supercalculateur Ruche commun à l'ENS Paris-Saclay et CentraleSupelec. 

\begin{figure}[!htb]
	\centering
	\includegraphics[scale=0.3]{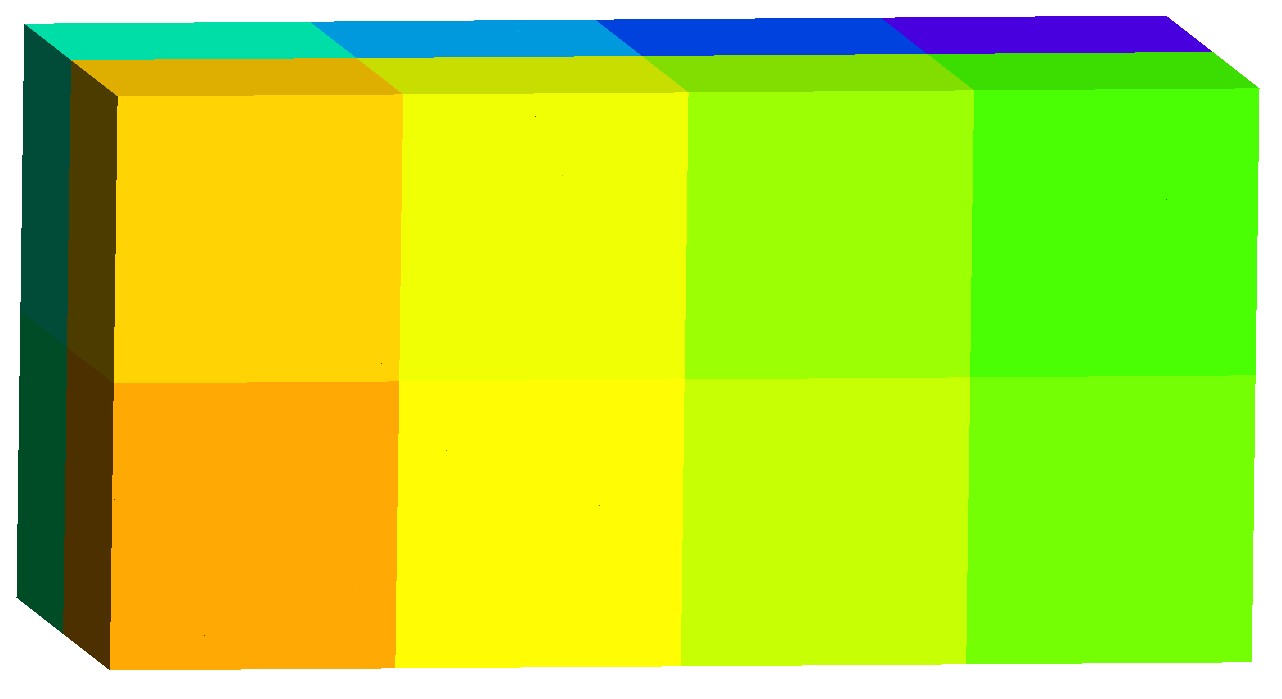}
	\caption{16 sous-domaines avec $\Omega^{0} = \emptyset$ }\label{fig:sds}
\end{figure}

Pour cette étude, nous avons retenu une géométrie à $2\times 2 \times 4$ patchs, voir la figure~\ref{fig:sds}. Le modèle global complet (16 cubes) comporte environ 8000 degrés de liberté, chaque modèle fin (1 cube avec inclusion) comporte 3750 degrés de liberté. 

%avons construit nos propres géometries, par exemple l'un des cas sur lequel on a testé est le suivant avec 16  sous-domaines, l'idèe est de passer d'une representation grossière avec un comportement homogène dans tous les sous-domaine, qu'on va venir corriger localement avec l'insertion d'une sphère dans laquelle on considère un comportement heterogène par rapport au reste du sous-domaine.\\
%Le code qu'on a utilisé a été developé en python et on se sert d'une bibliothèque élément finis "getfem cite" pour construire le modele à partir de la lecture de nos maillages realisés avec gmsh.
%Les resultas ont été realisé sur le supercalculateur Ruche commun à l'ens paris saclay et centrasupelec.\\
\begin{figure}[hbt]
	\centering
	\includegraphics[scale=1]{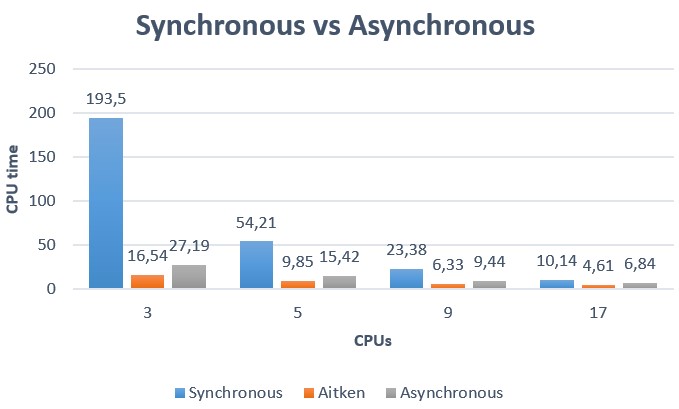}
	\caption{Comparaison des temps de calcul suivant les méthodes et ressources allouée}\label{fig:16sdtime}
\end{figure}

Notre étude consiste à faire varier les ressources allouées au calcul. Le tableau~\ref{tab:16sdit} présente les performances en termes d'itérations et la figure~\ref{fig:16sdtime} en termes de temps de calcul. Nous comparons les versions synchrone et asynchrone pour un même coefficient de relaxation déterminé de manière empirique pour optimiser les performances en asynchrone. Nous présentons également les performances de la méthode Aitken, qui est une variante synchrone où le coefficient de relaxation optimal est estimé à chaque itération par une heuristique faisant intervenir des calculs de produits scalaires. 

La méthode Aitken conduit aux meilleures performances pour tous nos indicateurs. Néanmoins, on observe que la méthode asynchrone n'est pas très éloignée en termes de temps de calcul.
\begin{table}[ht]\centering
	\begin{tabular}{|l||l|l|l||l|}\hline
		\textbf{CPU}  & Relaxation  &    \textbf{Synchrone}&    \textbf{Asynchrone} & \textbf{Aitken} \\  &$\omega$&\#it&\#itglo[\#itfin]&\#it (sync) \\\hline
		\textbf{3}   $\{1 + [8, 8 ]\}$ & 100  & 201   & 197[33 - 35]      		&16 \\  \hline
		\textbf{5}  $\{1 + [4, \cdots, 4 ]\}$ &  200 & 100  &   108 [34 - 41]     & 16     \\ \hline
		\textbf{9}   $\{1 + [2, \cdots,2 ]\}$   &300  &  67 &  61 [41 - 55] & 16 \\  \hline
		\textbf{17} $\{1 + [1, \cdots, 1]\}$ & 510 & 39 & 41 [54 - 75]&16    \\ \hline
	\end{tabular}\caption{Nombre d'itérations pour les différentes variantes, le coefficient de relaxation est optimisé pour les performances asynchrones. En asynchrone on distingue les itérations sur le modèle global et les patchs fins.}\label{tab:16sdit}
\end{table}
Dans la figure ci-dessus, nous pouvons voir le temps de calcul effectué par le modèle asynchrone
et celui effectué en synchrone. Nous commençons avec 8 sous-domaines par CPU + un CPU pour calculer le problème global, et en augmentant le nombre de CPUs à chaque fois, nous terminons avec un sous-domaine par CPU. Le résultat montre que globalement le temps de calcul du modèle asynchrone relaxé avec un coefficient constant se rapproche des temps de calcul du modèle synchrone avec L'Aitken qui est un algorithme d'optimisation de coefficient de relaxation d'une itèration à l'autre et qui converge en très peu d'itération, pour le meme coefficient de relaxation constant pour les deux modèles on peut voir que le modèle asynchrone est bien plus rapide.
Dans le tableau, nous résumons le nombre d'itérations effectuées dans chaque cas
en synchrone on fait le même nombre d'itérations partout contrairement à l'asynchrone
où chacun fait son propre nombre d'itérations qui diffèrent entre le global et les locaux.

\section{Conclusion}
Nous avons présenté une version asynchrone de la méthode de calcul global/local non-intrusive, et nous avons proposé une mise en œuvre par des techniques de parallélisation MPI RMA. Le cadre global/local est avantageux pour l'asynchronisme car on peut facilement prouver et contrôler la convergence.

Les performances présentées sont intéressantes : en terme de temps de calcul, la méthode asynchrone se rapproche de la redoutable variante Aitken du calcul synchrone. 

Durant la présentation, d'autres exemples académiques avec plus de sous-domaines et un exemple industriel seront présentés. D'autres architectures, moins défavorable qu'un cluster homogène, seront également testées.

	\medskip
	\noindent\textbf{Remerciements} : ce travail est réalisé dans le cadre du projet ANR ADOM [ANR-18-CE46-0008].
	% ----------------------------------------------------------------------
	
	% ----------------------------------------------------------------------
	

\begin{thebibliography}{1}
		% ----------------------------------------------------------------------
		\bibitem{Gendre}
		Lionel Gendre, Olivier Allix, Pierre Gosselet, and François Comte \emph{Non-intrusive and exact
			global/local techniques for structural problems with local plasticity.}, Computational Mechanics, 44(2):233-245, 2009.
		\bibitem{Hecht}
		F. Hecht, A. Lozinski, and O. Pironneau. \emph{Numerical zoom and the Schwarz algorithm.}, Proceedings of the 18th conference on domain decomposition methods, 2009.
		\bibitem{Magoules1}
		F Magoules, C Venet. \emph{Asynchronous iterative sub-structuring methods}, Mathematics and Computers in Simulation, 145, 34-49, 2018
		\bibitem{Garay}
		José C Garay, Frédéric Magoules, Daniel B Szyld. \emph{Synchronous and asynchronous optimized Schwarz method for Poisson's equation in rectangular domains}, Research Report 17-10-18, Department of Mathematics, Temple University, 2017.
		\bibitem{Magoules2}
		F. Magoules, D.B. Szyld, C. Venet. \emph{Asynchronous optimized Schwarz methods with and without overlap}, Numerische Mathematik,1,137, 199-227, 2017.
		\bibitem{Glusa}
		Christian Glusa, E. Boman, E. Chow, S. Rajamanickam, D. Szyld  \emph{Sacalable asynchronous domain decomposition solvers}, SIAM Journal on Scientific Computing, 42(6), 384–409, 2020.
		\bibitem{Guillaume} Guillaume Gbikpi-Benissan \emph{Méthodes asynchrones de décomposition de domaine pour le calcul massivement parallèle}, Thèse de l'école CentraleSupelec, 2017. 
		\bibitem{Gosselet}
		Pierre Gosselet, Maxime Blanchard, Olivier Allix, and Guillaume Guguin. \emph{Non-invasive globallocal
			coupling as a Schwarz domain decomposition method: acceleration and generalization},Advanced Modeling and Simulation in Engineering Sciences, 5(4), 2018.
		\bibitem{Allix}
		Olivier Allix, Pierre Gosselet \emph{Non Intrusive Global/Local Coupling Techniques in Solid Mechanics: An Introduction to Different Coupling Strategies and Acceleration Techniques},book : Modeling in Engineering Using Innovative Numerical Methods for Solids and Fluids. Springer, Cham, 203-220, 2020.
		\bibitem{Duval}
		Mickael Duval, Jean-Charles Passieux, Michel Salaun and Stéphane Guinard \emph{Non-intrusive
			coupling: recent advances and scalable nonlinear domain decomposition.}, Archives of Compu-
		tational Methods in Engineering, 1-22, 2014.
		\bibitem{Magoules3}
		Frédéric Magoulès, Guillaume Gbikpi-Benissan \emph{Non-intrusive
			coupling: recent advances and scalable nonlinear domain decomposition.}, Advances in Engineering Software, Elsevier, 116-133, 2018.
		\bibitem{Yamazaki}
		Ichitaro Yamazaki, Edmond Chow, Aurelien Bouteiller, Jack Dongarra \emph{Performance of asynchronous optimized Schwarz with one-sided communication},  Parallel Computing, 86, 66-81, 2019.
		\bibitem{Eisner}L. Eisner, I. Koltracht, and M. Neumann. \emph{Convergence of sequential and asynchronous nonlinear paracontractions.}, Numerische Mathematik, 62:305, 319, 1992.
		\bibitem{Renard}
		Yves Renard, Konstantinos Poulios \emph{GetFEM: Automated FE modeling of multiphysics problems based on a generic weak form language.}, 2020.
		\bibitem{Geuzaine}
		Christophe Geuzaine, Jean-François Remacle \emph{Gmsh: a three-dimensional nite element mesh generator with built-in pre- and post-processing facilities.},International Journal for Numerical Methods in Engineering, 0:1-24, 2009.
		\bibitem{Magoules4}
		Frédéric Magoulès, Guillaume Gbikpi-Benissan \emph{Distributed Convergence Detection Based
			on Global Residual Error Under Asynchronous Iterations}, IEEE transactions on parallel and distributed systems, 29, 2018.
		\bibitem{Miellou}
		J.C.Miellou, P.Spiteri, D.El Baz \emph{A new stopping criterion for linear perturbed
			asynchronous iterations.},Journal of Computational and Applied Mathematics 219, 471 – 483,  2008
		%\bibitem{acte} P. Auteur. \emph{Titre de l'acte}, Titre de l'ouvrage, Éditeur, page1-pageN, Année.
		
		
		% ----------------------------------------------------------------------
	\end{thebibliography}
\end{document}